\documentclass[runningheads]{llncs}
\usepackage{comment, graphicx, booktabs, hyperref, subcaption, amsmath, amssymb, color, xspace, siunitx, multirow}
\usepackage[inline]{enumitem}
\graphicspath{{figure/}}

\begin{document}
\title{Less is More: Culling the Training Set to Improve Robustness of Deep Neural Networks}
\titlerunning{Less is More}
\author{Yongshuai Liu\thanks{Equal contribution} \and
Jiyu Chen\inst{*} \and
Hao Chen}
\authorrunning{Y. Liu et al.}
\institute{University of California, Davis\\ \email{\{yshliu, jiych, chen\}@ucdavis.edu}}
\maketitle              

\begin{abstract}

	Deep neural networks are vulnerable to adversarial examples. Prior defenses attempted to make deep networks more robust by either changing the network architecture or augmenting the training set with adversarial examples, but both have inherent limitations. Motivated by recent research that shows outliers in the training set have a high negative influence on the trained model, we studied the relationship between model robustness and the quality of the training set. We first show that outliers give the model better generalization ability but weaker robustness. Next, we propose an adversarial example detection framework, in which we design two methods for removing outliers from training set to obtain the sanitized model and then detect adversarial example by calculating the difference of outputs between the original and the sanitized model. We evaluated the framework on both MNIST and SVHN. Based on the difference measured by Kullback-Leibler divergence, we could detect adversarial examples with accuracy between 94.67\% to 99.89\%.
	
\end{abstract}

\section{Introduction}
\label{sec:Introduction}

Deep neural networks have demonstrated impressive performance on many hard perception problems~\cite{Krizhevsky:2012,Lecun:1998}. However, they are vulnerable to adversarial examples~\cite{Szegedy:2014,Goodfellow:2015,Moosavi:2016}, which are maliciously crafted to be perceptually close to normal examples but which cause misclassification. Prior defenses against adversarial examples fall into the following categories: 
\begin{enumerate*}
	\item Incorporating adversarial examples in the training set, a.k.a. adversarial training~\cite{Szegedy:2014,Goodfellow:2015}
	\item Modifying the network architecture or training method, e.g., defense distillation~\cite{Papernot:2016:Distillation}
	\item Modifying the test examples, e.g., MagNet~\cite{Meng:2017:MagNet}
\end{enumerate*}
The first defense requires knowledge about the process for generating adversarial examples, while the last two defenses require high expertise and are often not robust~\cite{Carlini:2017}.

We propose a new direction to strengthen deep neural networks against adversarial examples. Recent research showed that outliers in the training set are highly influential on the trained model. For example, outliers may be ambiguous images on which the model has low confidence and thus high loss~\cite{Koh:2017}. Our insight is that outliers give the model better generalization ability however also make the model more sensitive to adversarial examples. When we detect and discard outliers in the training set, the new model will be less sensitive to adversarial examples. And we utilize the sensitivity difference between the original model and the new model to distinguish adversarial examples from normal examples.

We call the process of removing outliers from the training set \emph{sanitization}.\footnote{Unlike \emph{data sanitization}, which commonly modifies individual datum, we modify no example but merely remove outliers from the training set.} We propose two methods for detecting outliers. First, for some AI tasks, we may find canonical examples. For example, for handwritten digit classification, we may use computer fonts as canonical examples. We trained a \emph{canonical model} using canonical examples, and then used the canonical model to detect outliers in the training set. We call this method \emph{canonical sanitization}. Second, for AI tasks without canonical examples, we considered examples with large training errors as outliers. We call this method \emph{self sanitization}.

After culling the training set, we trained a model called the \emph{sanitized model}. We compared the robustness of the unsanitized model, which was trained on the entire training set, with the sanitized model on adversarial examples using two criteria with respect to different attack methods. For IGSM attack, the criterion is classification accuracy of adversarial examples. For Carlini \& Wagner attack, the criterion is the average distortion. In \autoref{sec:classification}, the result of sanitization exactly validates that the outliers help model do better generalization meanwhile decrease the robustness. Given the result, the sanitized models allow us to detect adversarial examples which is shown in \autoref{sec:detection}.

To measure the sensitivity difference, we computed the Kullback-Leibler divergence from the output of an example on the unsanitized model to the output of the same example on the sanitized model and found that this divergence was much larger for adversarial examples than for normal examples. Based on this difference, we were able to detect the adversarial examples generated by the Carlini \& Wagner attack on MNIST and SVHN at 99.26\% and 94.67\% accuracy, respectively. Compared to prior work for detecting adversarial examples (e.g., \cite{Metzen:2017}), this approach requires no knowledge of adversarial examples.

We make the following contributions.

\begin{itemize}
	\item We propose two methods for detecting outliers in the training set: canonical sanitization and self-sanitization. By performing data sanitization, we show how the outliers will affect the model's robustness and generalization ability.
	
	\item We propose a new adversarial example detection framework based on the sanitized model. The detector leverages the Kullback-Leibler divergence from the unsanitized model to the sanitized model. Neither modifications to the model structure nor data preprocessing methods are required.
\end{itemize}

\section{Methodology}
\label{sec:methodology}

\subsection{Definitions}
\label{sec:definitions}

\begin{itemize}
	
	\item \textit{Normal examples} are sampled from the natural data generating process. For examples, images of handwritten digits.
	
	\item \textit{Outliers} are examples in the training set of normal examples. They are difficult to classify by humans. \emph{Sanitization} is the process of removing outliers from the training set.
	
	\item \textit{Adversarial examples} are crafted by attackers that are perceptually close to normal examples but that cause misclassification.
	
	\item \textit{Unsanitized models} are trained with all the examples in the training set. We assume that the training set contains only normal examples.
	
	\item \textit{Sanitized models} are trained with the remaining examples after we remove outliers from the training set.
	
\end{itemize}

\subsection{Sanitization}

Sanitization is the process of removing outliers from the training set. We propose two automatic sanitization methods.

\paragraph{\textbf{Canonical sanitization}}

This approach applies to the AI tasks that have canonical examples. For example, for handwritten digit, computer fonts may be considered canonical examples\footnote{Some computer fonts are difficult to recognize and therefore are excluded from our evaluation}. Based on this observation, we use canonical examples to discard outliers in our training set $\mathbb{X}$ by the following steps:

\begin{itemize}
	
	\item Augment the set of canonical examples by applying common transformations, e.g., rotating and scaling computer fonts.
	
	\item Train a model $f$ using the augmented canonical examples.
	
	\item Use $f$ to detect and discard outliers in the training set
	$\mathbb{X}$. An example $\boldsymbol{x}^{(i)}$ is an outlier if
	$f(\boldsymbol{x}^{(i)})$ has a low confidence on $y^{(i)}$, the class
	for $\boldsymbol{x}^{(i)}$.
	
\end{itemize}

\paragraph{\textbf{Self sanitization}}

Not all AI tasks have canonical examples. For such tasks, we use all the examples to train a model, and then discard examples that have high training errors.

\paragraph{}
After removing outliers from the original training set, we get a sanitized set which is used to train a model, called \emph{sanitized model}. Then, we evaluate if the sanitized model is more robust than unsanitized models using two metrics: classification accuracy and distortion of adversarial examples.

\subsection{Detecting adversarial examples}
\label{sec:kld}
We take advantage of the Kullback-Leibler divergence~\cite{Kullback:1951} between the outputs of the original and the sanitized models to depict the difference of sensitivity to the adversarial examples. The Kullback-Leibler divergence from a distribution $P$ to $Q$ is defined as

\[
D_{\mathrm{KL}}\left ( P \; \middle\| \; Q \right ) = \sum_i P\left ( i \right )\log\frac{P\left ( i \right )}{Q\left ( i \right )}
\]

By setting a proper threshold, we are able to detect nearly all adversarial examples with acceptable false reject rate. No modifications to the original model structure or other data outside the original dataset are required.

The detection method is hard to distinguish between adversarial examples and normal examples when the distortion of the adversarial image is very small. To address this problem, we designed a complete adversarial example detection framework. We will discuss the framework detailedly in \autoref{sec:detection}.

\section{Evaluation}
\label{sec:evaluation}

\subsection{Set up}

We used two data sets, MNIST\footnote{http://yann.lecun.com/exdb/mnist/} and SVHN\footnote{http://ufldl.stanford.edu/housenumbers/}, to evaluate our proposed method. We performed both canonical sanitization and self sanitization on MNIST and only self sanitization on SVHN. For SVHN, we pre-processed it with the following steps to get individual clean digit images. After the process, we obtained \num{40556} images from the original SVHN training set and \num{9790} test images from the original SVHN test set.

\begin{enumerate}
	\item Cropping individual digits using the bounding boxes.
	\item Discarding images whose either dimension is less than 10.
	\item Resizing the larger dimension of each image to 28 while keeping the aspect ratio, and then padding the image to $28 \times 28$. When padding an image, we used the average color of the border as the padding color.
\end{enumerate}

The models we used to train these two datasets are different. We designed Convolutional Neural Networks for MNIST and SVHN separately. Correspondingly, we achieved an accuracy of $99.3\%$ and $98.62\%$ on the \emph{unsanitized models}.

\begin{itemize}
	
	\item MNIST CNN: Input $\to$ (Conv + Pool) * 2 $\to$ FC $\to$ FC  $\to$ Output
	
	\item SVHN CNN: Input $\to$ (Conv + Conv + Pool) * 3 $\to$ FC $\to$ FC  $\to$ Output
	
\end{itemize}

Given the trained model, we performed two popular attacks, Iterative Gradient Sign Method (IGSM)~\cite{Goodfellow:2015,Kurakin:2016} and Carlini \& Wagner's attack~\cite{Carlini:2017}, to attack the CNN models. We will discuss the attacks in \autoref{sec:classification}.

\subsection{Sanitization}

\paragraph{\textbf{Canonical sanitization}}
\label{sec:canonical}

We did canonical sanitization on MNIST by discarding outliers that are far different from canonical examples. We chose 340 fonts containing digits as canonical examples. To accommodate variations in handwriting, we also augmented the fonts by scaling and rotation. After the augmentation, we acquired $\num{71400}$ images, from which we randomly chose 80\% as the training set and the remaining 20\% as the test set. We trained the MNIST CNN on canonical examples and achieved an accuracy of 98.7\%. We call this the \textit{canonical model}.

\begin{figure}[t]
	\begin{center}
		\begin{subfigure}{0.35\linewidth}
			\centering \includegraphics[width=0.5\linewidth]{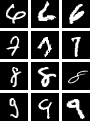}
			\caption{Confidence $<0.7$}
			\label{fig:low_confidence}
			\vspace{-0.15in}
		\end{subfigure}
		\begin{subfigure}{0.35\linewidth}
			\centering \includegraphics[width=0.5\linewidth]{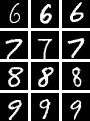}
			\caption{Confidence $>0.999\,99$}
			\label{fig:high_confidence}
			\vspace{-0.15in}
		\end{subfigure}
	\end{center}
	
	\caption{Examples in MNIST with low and high confidence, respectively.}
	\label{fig:confidence}
	\vspace{-0.1in}
\end{figure}

We fed each example in MNIST training set to the canonical model. If the example's confidence score of the correct class was below a threshold, we considered it an outlier and discarded it. \autoref{fig:confidence} shows examples with low and high confidence. \autoref{tbl:font-set} shows the number of examples left under different thresholds. We used these examples to train the \textit{sanitized models}.

\begin{table}[t]
	\centering
	\begin{tabular}{lrrrrrrr}
		\toprule
		Threshold     &    0 & 0.7 & 0.8 & 0.9 & 0.99 & 0.999 & \num{0.9999} \\
		Set size  &   \num{60000}  & \num{51241} & \num{50425} & \num{49128} & \num{44448} & \num{38230} & \num{30618} \\
		\bottomrule
	\end{tabular}
	\vspace{0.1in}
	\caption{Size of the MNIST training set after discarding examples whose confidence scores on the canonical model are below a threshold}
	\label{tbl:font-set}
	\vspace{-0.35in}
\end{table}

\paragraph{\textbf{Self-sanitization}}
\label{sec:self}
We did self sanitization on both MNIST and SVHN. To discard outliers in self sanitization, we trained the CNN for MNIST and SVHN separately, used the models to test every example in the training set, and considered examples whose confidence scores were below a threshold as outliers. \autoref{tbl:self-set-mnist} and \autoref{tbl:self-set-svhn} show the number of examples left under different thresholds. We used these examples to train the \textit{sanitized models}. \autoref{tbl:self-set-svhn} also shows that the sanitized models maintain high classification accuracy when it has adequate training data to prevent overfitting.

\begin{table}[t]
	\centering
	\caption{Size of the MNIST training set after discarding examples whose confidence scores on the self-trained model are below a threshold}
	\label{tbl:self-set-mnist}
	\centering
	\begin{tabular}{@{}lrrrrrr@{}}
		\toprule
		Threshold      &   0 & 0.999 & \num{0.9999} & \num{0.99999} & \num{0.999999} & \num{0.9999999} \\
		Set size&   \num{60000}  & \num{56435} & \num{52417} & \num{45769} & \num{36328} & \num{24678} \\
		\bottomrule
	\end{tabular}
	\vspace{-0.15in}
\end{table}

\begin{table}[t]
	\centering
	\caption{The sizes of the sanitized SVHN training set and the classification accuracy of the self-sanitized models at different thresholds}
	\label{tbl:self-set-svhn}
	\centering
	\begin{tabular}{SSS}
		\toprule
		\multicolumn{1}{c}{Threshold} & \multicolumn{1}{c}{Training set size} & \multicolumn{1}{c}{Classification accuracy (\%)}\\
		\midrule
		0 & 40556 & 94.26\\
		0.7 & 39330 & 93.68 \\
		0.8 & 38929 & 93.22  \\
		0.9 & 38153 & 92.74 \\
		0.99 & 34408 & 91.30  \\
		0.999 & 28420 & 89.41\\
		\bottomrule
	\end{tabular}
	\vspace{-0.15in}
\end{table}

\subsection{Robustness against adversarial examples}
\label{sec:classification}

We ran the IGSM and Carlini \& Wagner attacks on both the unsanitized and sanitized models.

\paragraph{\textbf{IGSM attack}}

\begin{figure}[t]
	\begin{center}
		\begin{subfigure}{0.49\linewidth}
			\centering \includegraphics[width=1\linewidth]{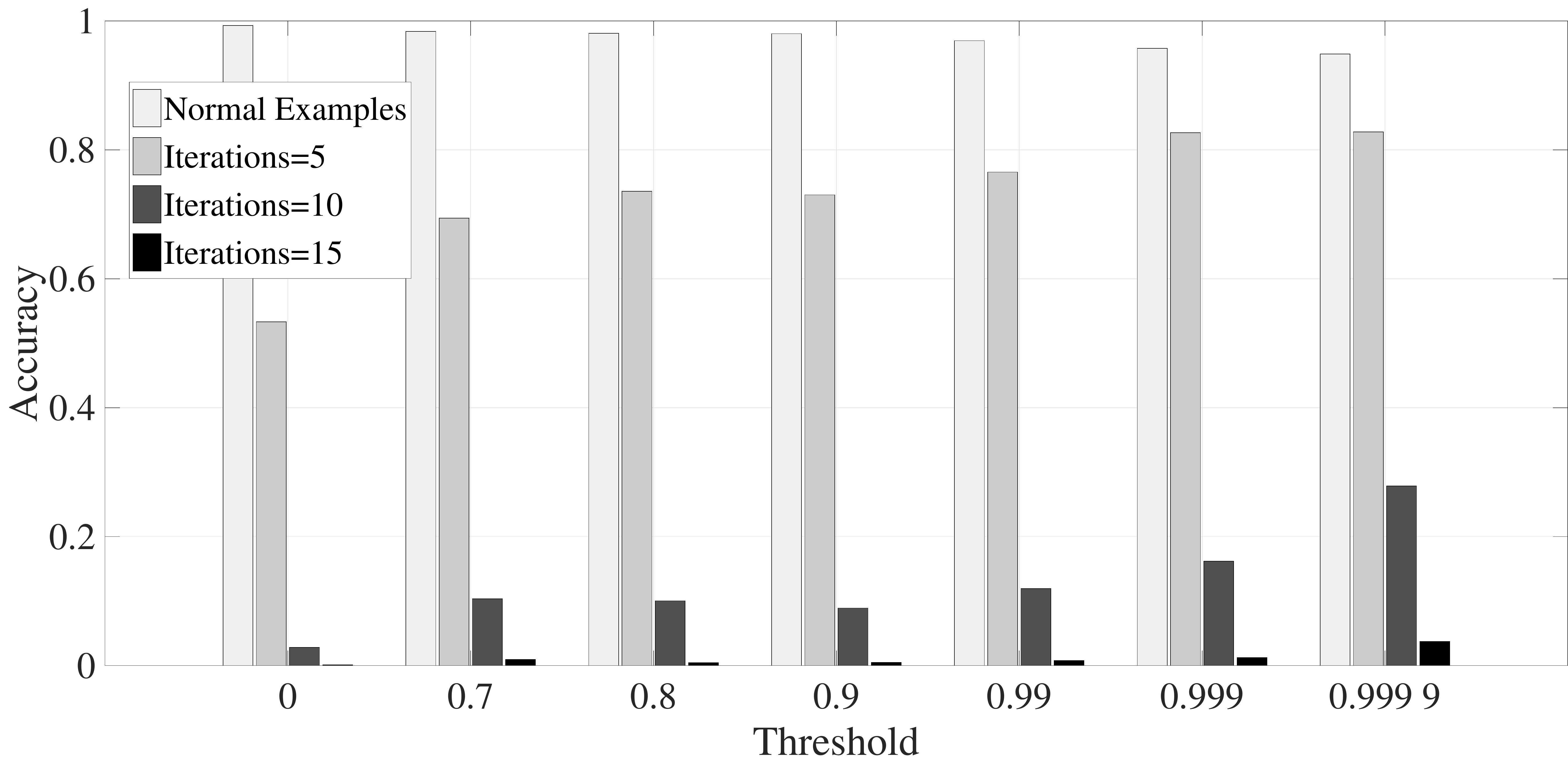}
			\caption{Canonical sanitization}		   			  
			\label{fig:font_fgsm}
			\vspace{-0.15in}
		\end{subfigure}
		\begin{subfigure}{0.49\linewidth}
			\centering \includegraphics[width=1\linewidth]{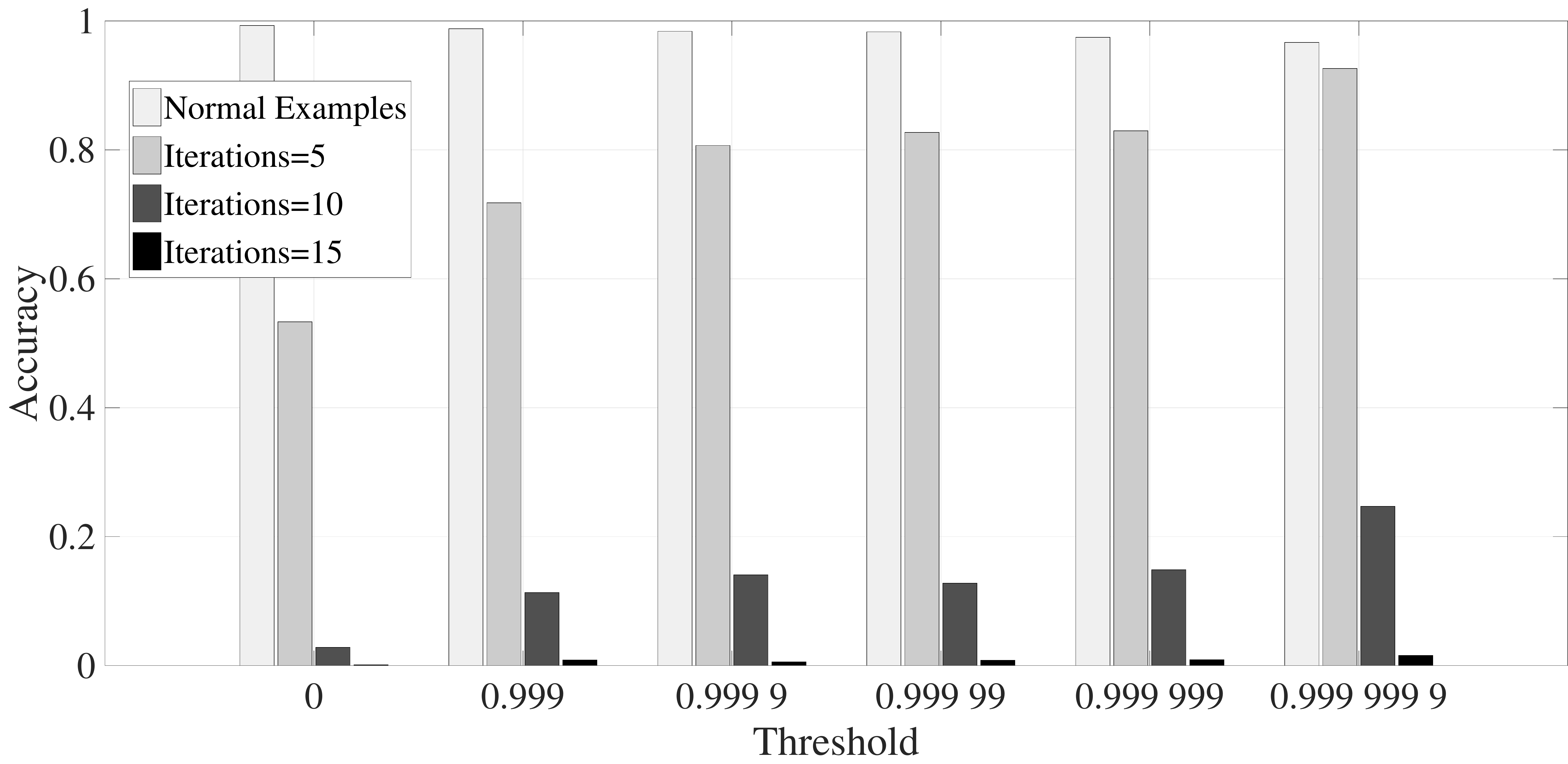}
			\caption{Self sanitization}		   			  
			\label{fig:self_fgsm}
			\vspace{-0.15in}
		\end{subfigure}
	\end{center}
	\caption{Classification accuracy of normal and IGSM adversarial MNIST examples in different threshold. The threshold 0 represents the original data set.}
	\label{fig:fgsm}
	\vspace{-0.1in}
\end{figure}

\autoref{fig:fgsm} compares the classification accuracy of the unsanitized and sanitized models on the adversarial examples generated by the IGSM attack on MNIST, where \autoref{fig:font_fgsm} and \autoref{fig:self_fgsm} correspond to canonical sanitization and self sanitization, respectively.

\autoref{fig:fgsm} shows that a higher threshold of sanitization increases the robustness of the model against adversarial examples and maintains classfication accuracy on normal examples. For example, on adversarial examples generated after five iterations of IGSM, the classification accuracy is 82.8\% with a threshold of \num{0.9999} in canonical sanitization, and is above 92.6\% with a threshold of \num{0.9999999} in self sanitization. For normal examples, the classfication accuracy is always higher than 95.0\% in different threshold.

\paragraph{\textbf{Carlini \& Wagner's attack}}

\begin{figure}[t]
	\begin{center}
		\begin{subfigure}{0.49\linewidth}
			\centering \includegraphics[width=1\linewidth]{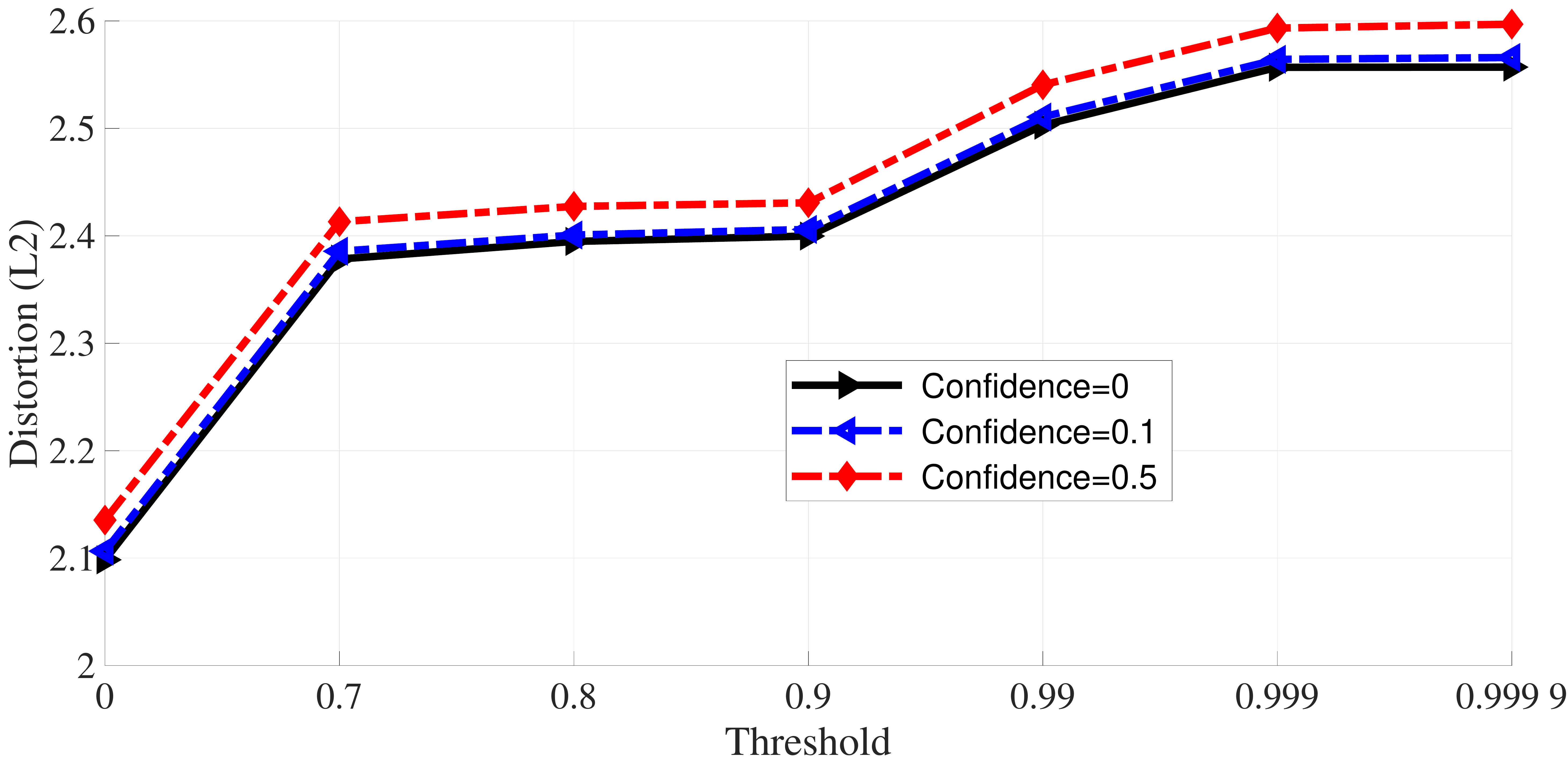}
			\caption{Canonical sanitization}		   			  
			\label{fig:font_carlini}
			\vspace{-0.15in}
		\end{subfigure}
		\begin{subfigure}{0.49\linewidth}
			\centering \includegraphics[width=1\linewidth]{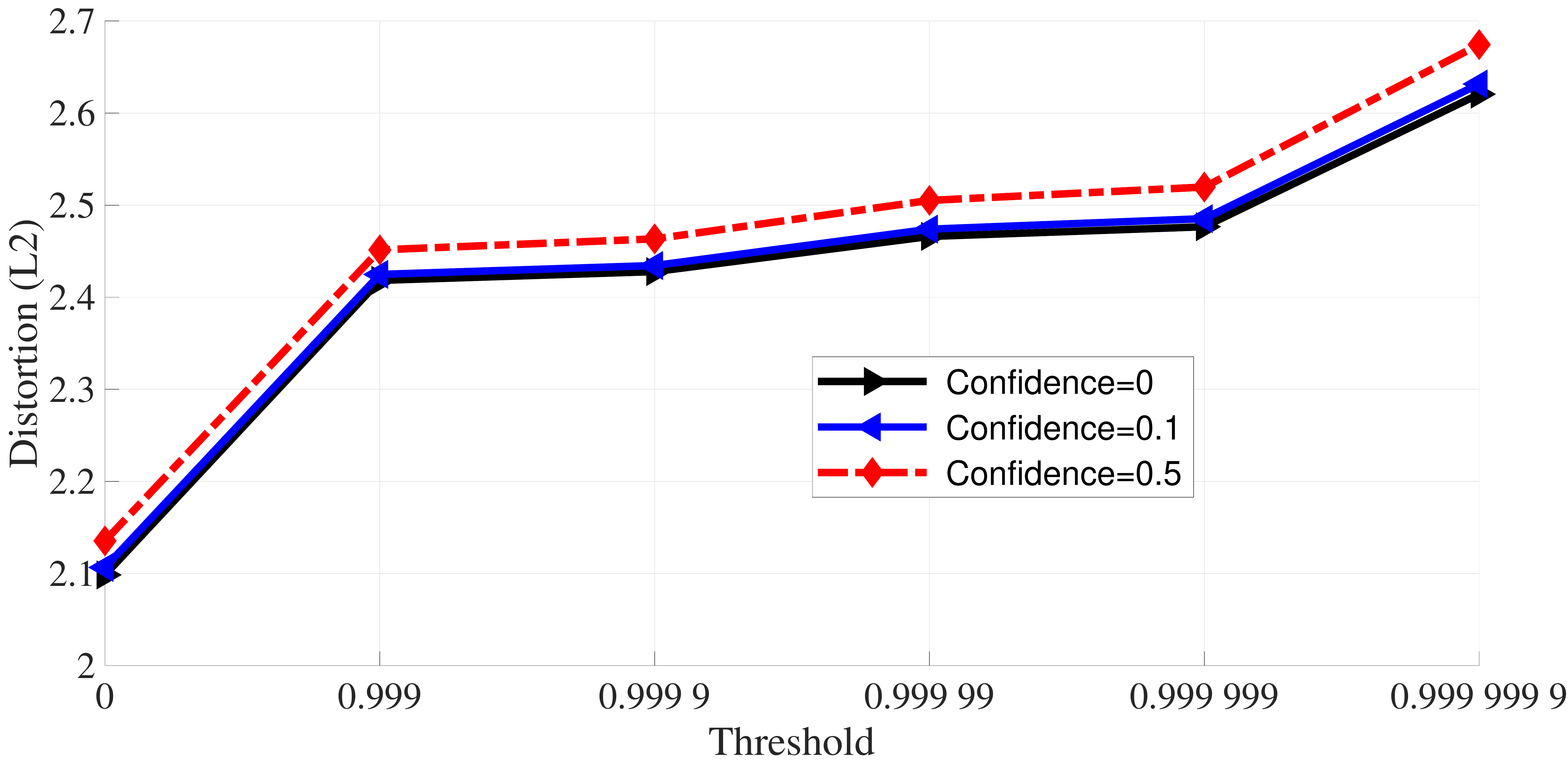}
			\caption{Self sanitization}		   			  
			\label{fig:self_carlini}
			\vspace{-0.15in}
		\end{subfigure}
	\end{center}
	\caption{Average $L_2$ distortions of normal and C\&W's adversarial MNIST examples in different threshold. The threshold 0 represents the unsanitized model.}
	\label{fig:carlini_mnist}
	\vspace{-0.25in}
\end{figure}

\begin{figure}[t]
	\begin{center}
		\begin{minipage}{0.49\linewidth}
			\centering \includegraphics[width=1\linewidth]{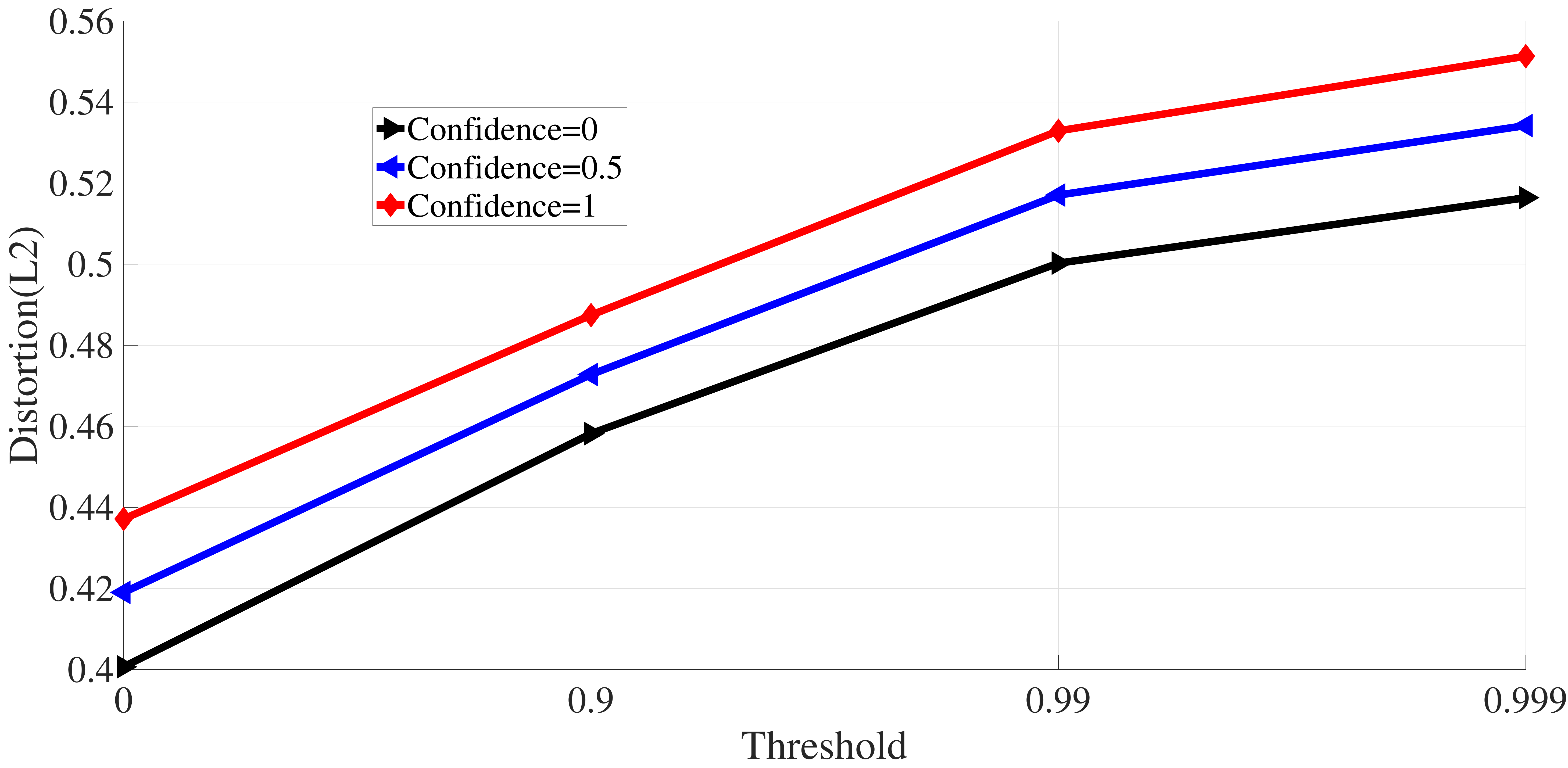}
			\caption{Average $L_2$ distortions of normal and C\&W's adversarial SVHN examples in different threshold. The threshold 0 represents the unsanitized model.}		   			  
			\label{fig:self_carlini_svhn}
			\vspace{-0.15in}
		\end{minipage}
		\begin{minipage}{0.49\linewidth}
			\centering \includegraphics[width=1\linewidth]{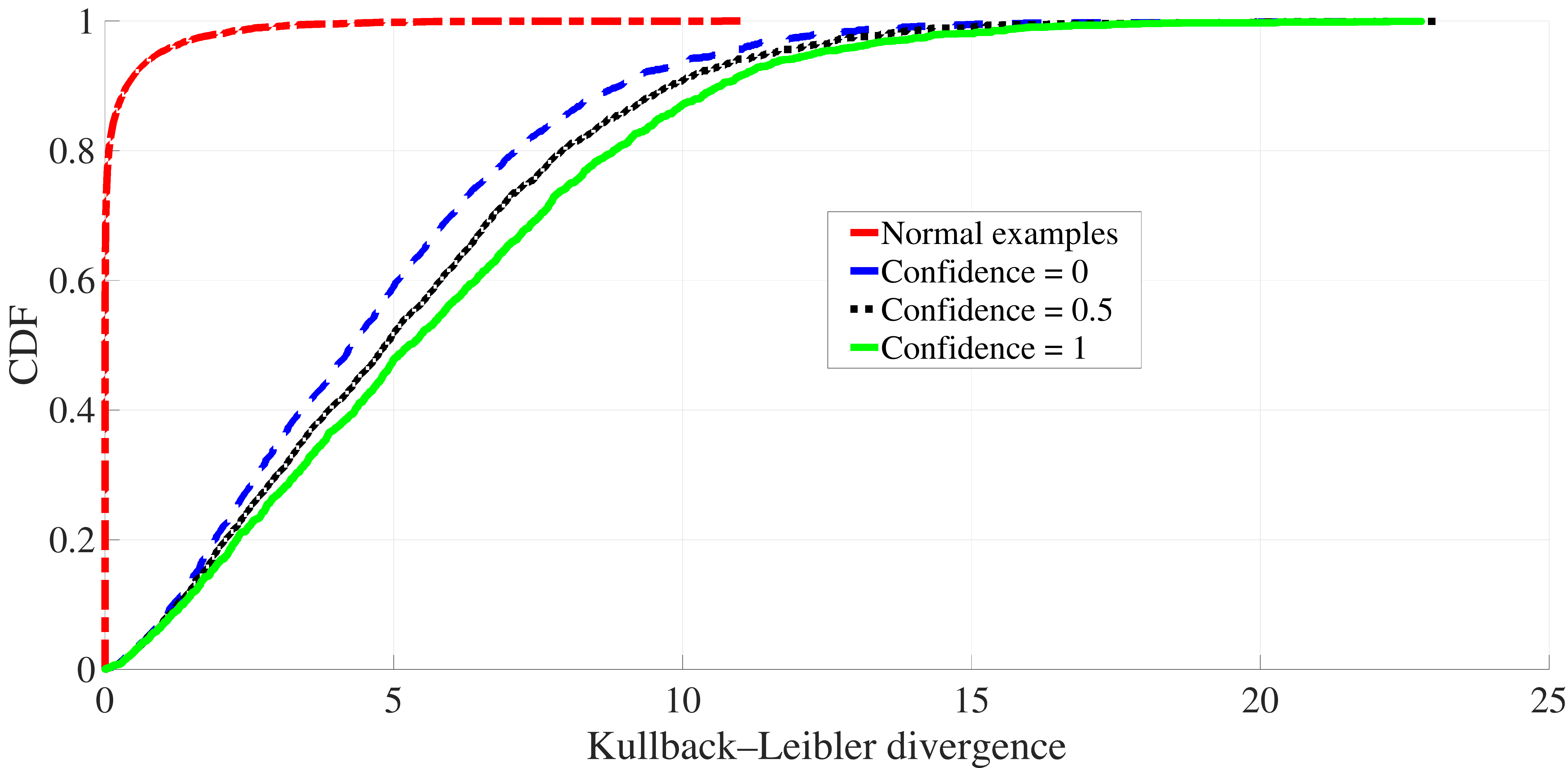}
			\caption{SVHN: CDF of the KL divergence from the output of unsanitized model to that of a sanitized model with different C\&W attack condidences. }		   			  
			\label{fig:dkl_carlini_self_svhn}
			\vspace{-0.15in}
		\end{minipage}
	\end{center}
	\vspace{-0.1in}
\end{figure}

We ran Carlini \& Wagner's $L_2$ target attack to generate adversarial examples on our sanitized models for both MNIST and SVHN. \autoref{fig:carlini_mnist} and \autoref{fig:self_carlini_svhn} show that the sanitized models forced the adversarial examples to add larger distortions in order to fool the sanitized models. The higher the threshold, the larger the distortion.

\paragraph{\textbf{Discussion}}
From the experiments, we concluded that the original dataset with outliers has higher generalization ability but weaker robustness. With the sanitization, the model obtained much more robustness by only sacrificing limited generalization ability.

\subsection{Detecting adversarial examples}
\label{sec:detection}

We evaluated the effectiveness of using the Kullback-Leibler divergence to detect adversarial examples (\autoref{sec:kld}).

\paragraph{\textbf{MNIST}}
We generated adversarial examples on two sanitized models on MNIST:

\begin{itemize}
	\item A canonical sanitized model. The discard threshold was set to be \num{0.9999}.
	
	\item A self sanitized model. The discard threshold was set to be \num{0.9999999}.
\end{itemize}

We computed the Kullback-Leibler divergence from the output of the unsanitized model to that of each of the sanitized models.

\begin{figure}[t]
	\begin{center}
		\begin{subfigure}{0.49\linewidth}
			\centering \includegraphics[width=1\linewidth]{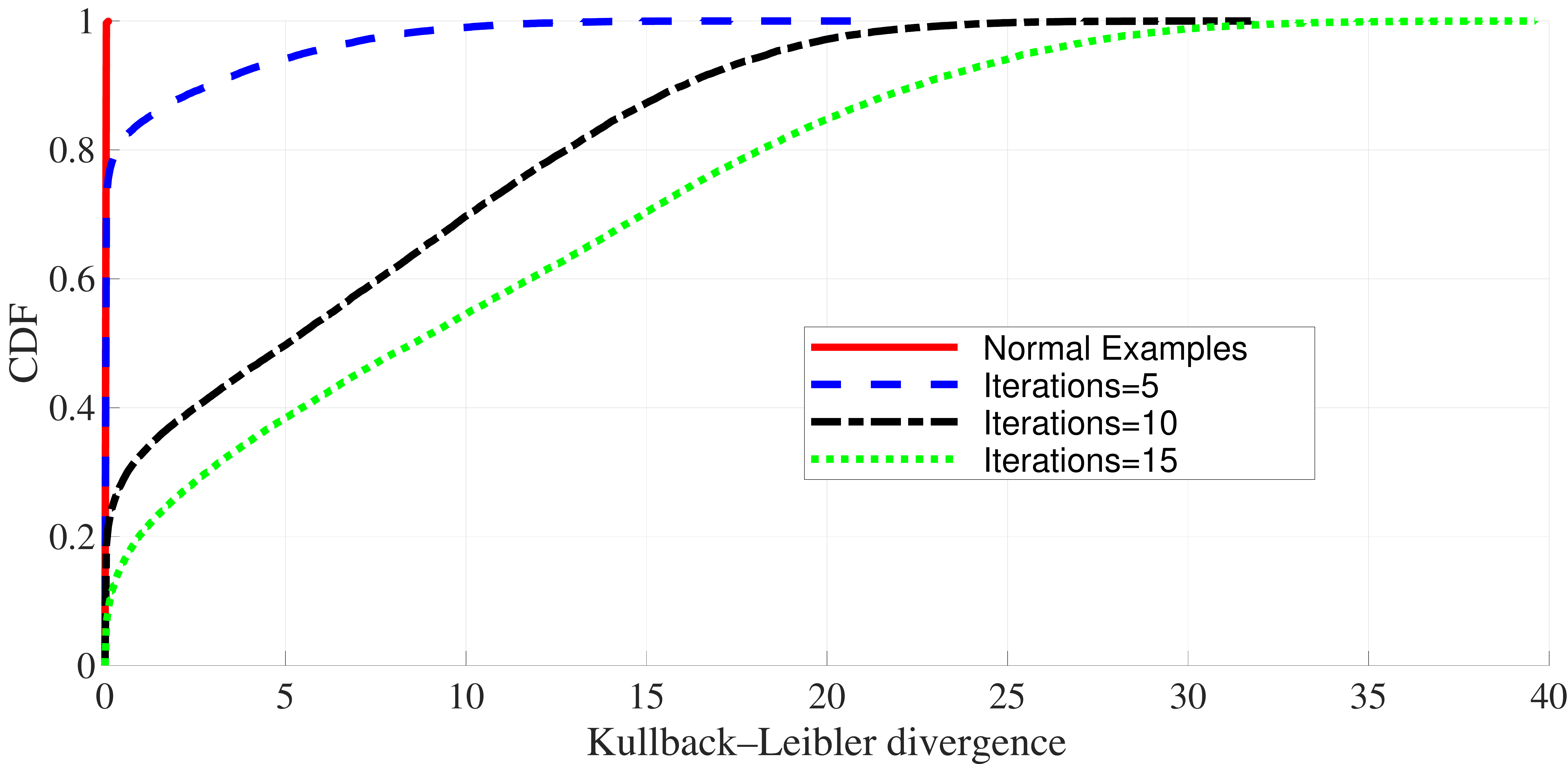}
			\caption{KL divergence from unsanitized model to canonical sanitized models}
			\label{fig:dkl_fgsm_font_cdf}
			\vspace{-0.15in}
		\end{subfigure}
		\begin{subfigure}{0.49\linewidth}
			\centering \includegraphics[width=1\linewidth]{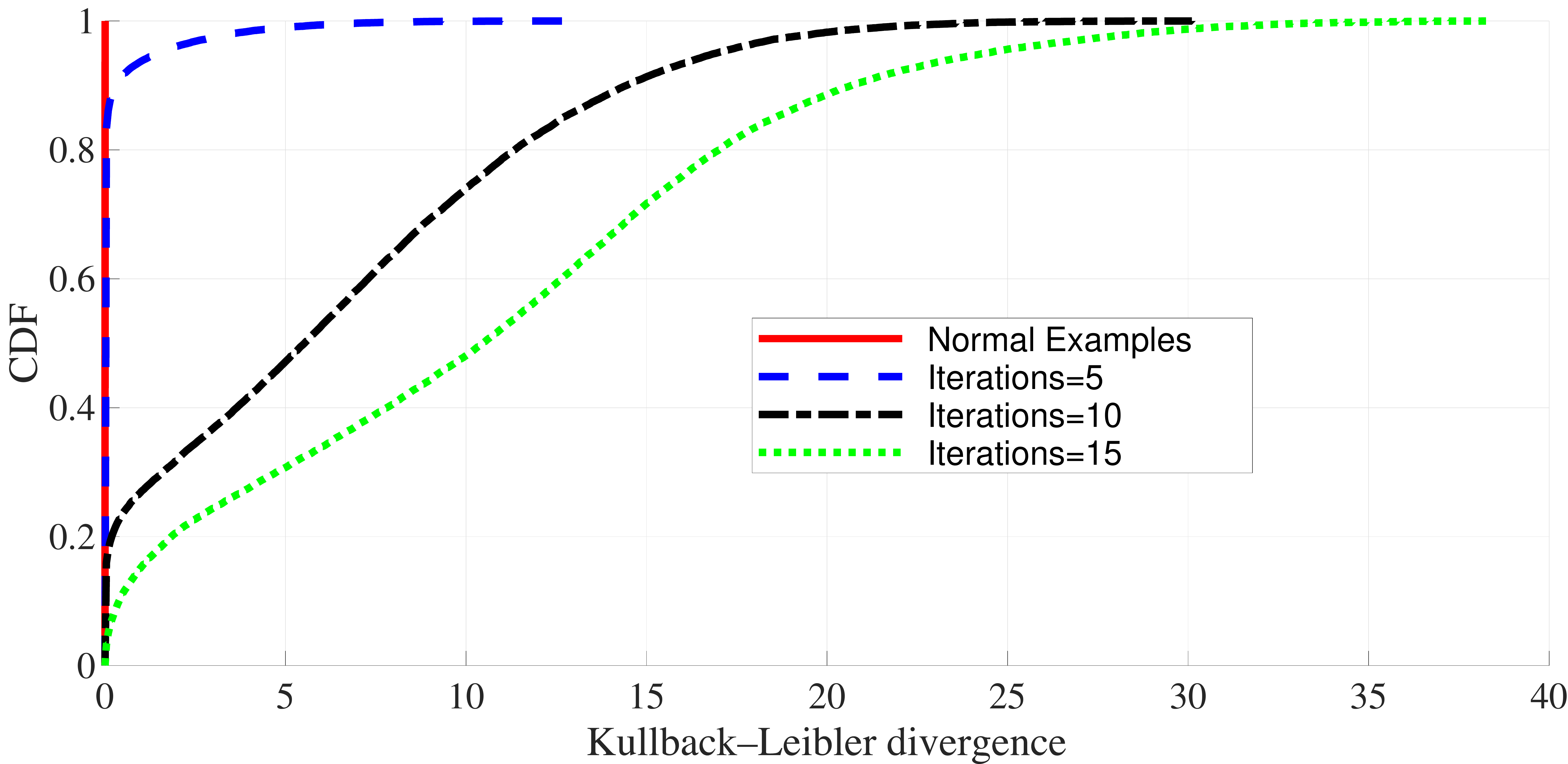}
			\caption{KL divergence from unsanitized model to self sanitized models}
			\label{fig:dkl_fgsm_self_cdf}
			\vspace{-0.15in}
		\end{subfigure}
	\end{center}
	\caption{MNIST: CDF of KL divergence from the output of the unsanitized model to the output of a sanitized model with different IGSM iterations.}
	\label{fig:dkl_fgsm}
	\vspace{-0.2in}
\end{figure}

\begin{figure}[t]
	\begin{center}
		\begin{subfigure}{0.49\linewidth}
			\centering \includegraphics[width=1\linewidth]{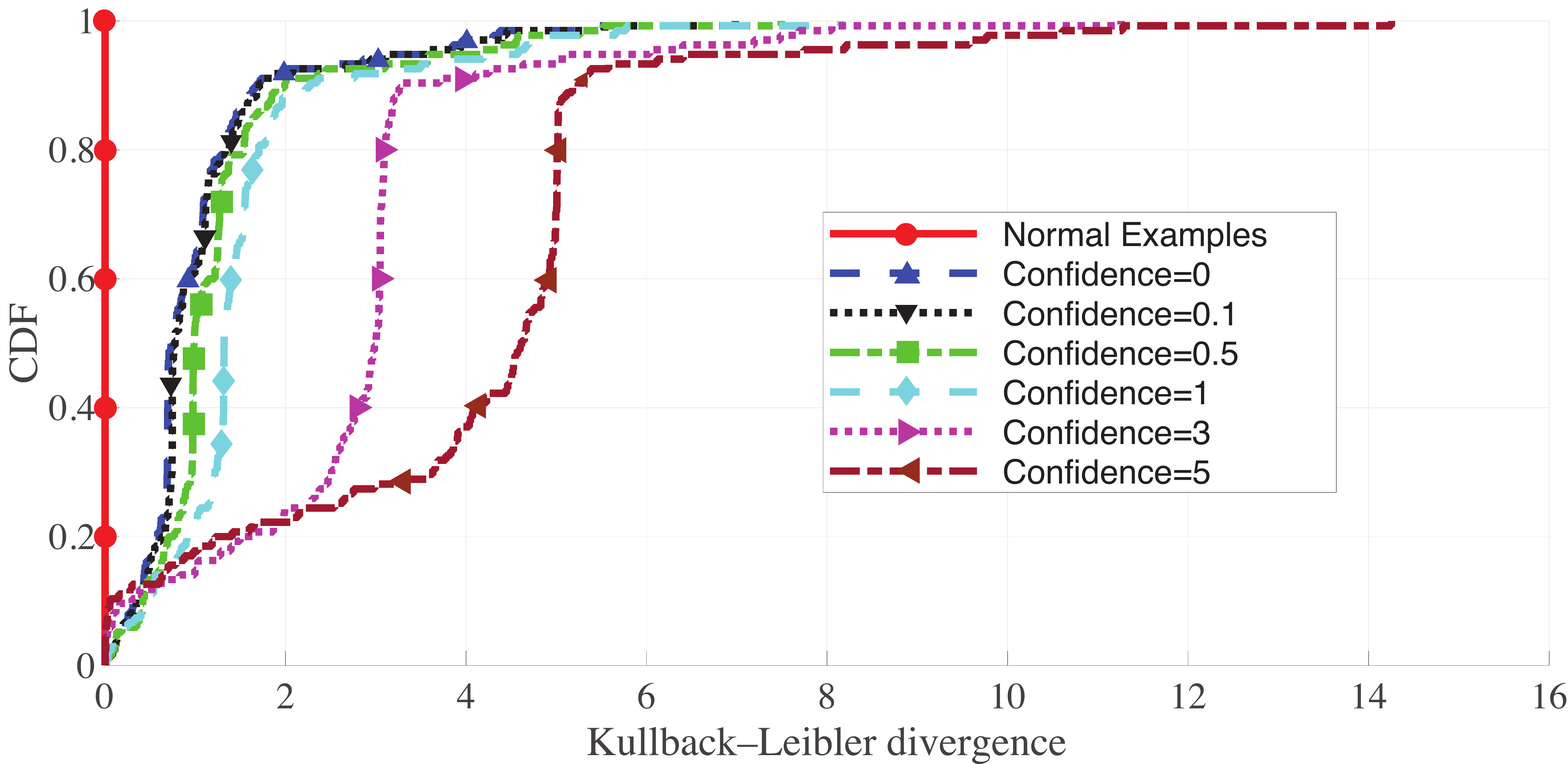}
			\caption{KL divergence from the unsanitized model to canonical sanitized models}
			\label{fig:dkl_carlini_font_cdf}
			\vspace{-0.15in}
		\end{subfigure}
		\begin{subfigure}{0.49\linewidth}
			\centering \includegraphics[width=1\linewidth]{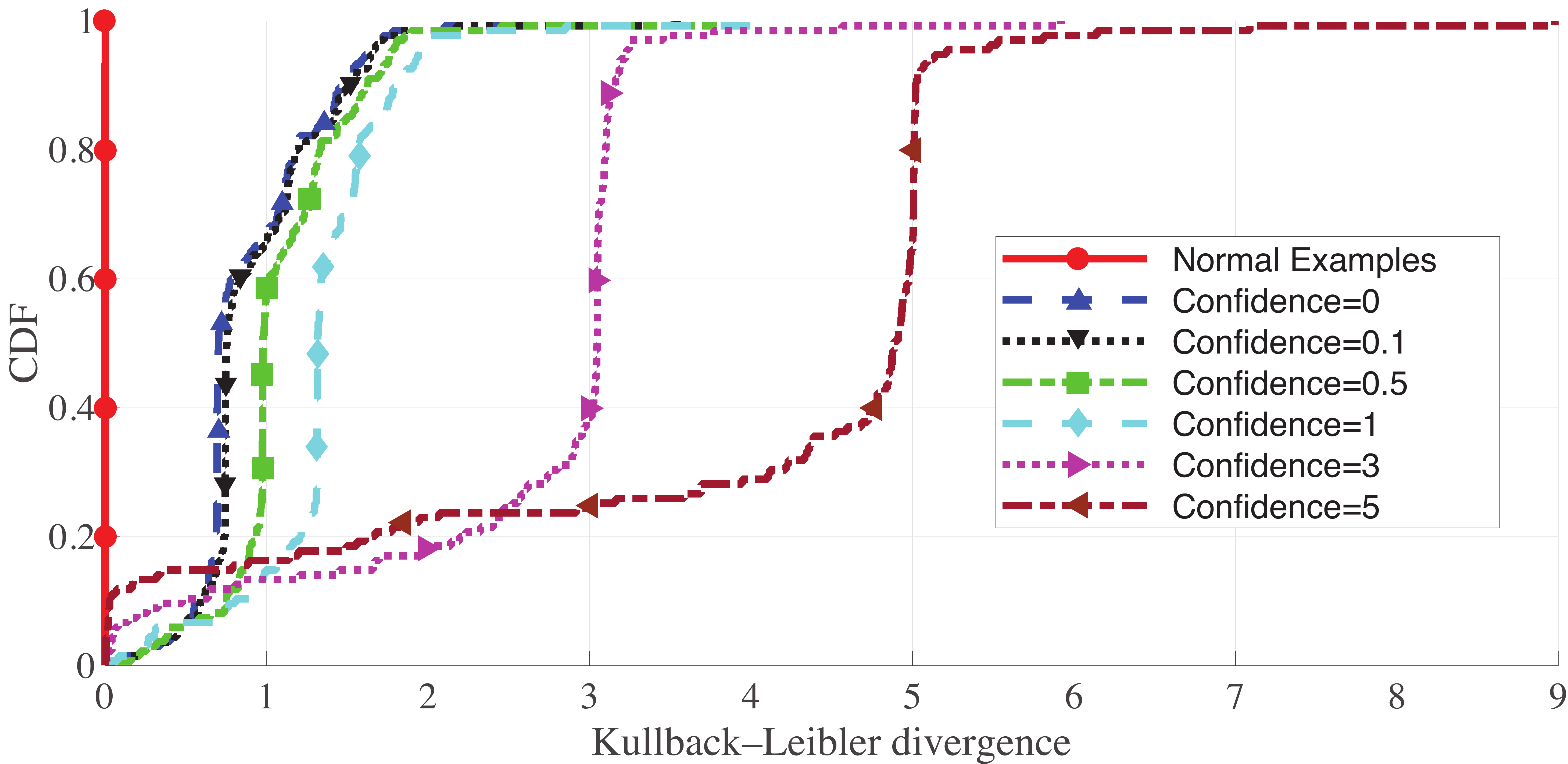}
			\caption{KL divergence from unsanitized model to self sanitized models}
			\label{fig:dkl_carlini_self_cdf}
			\vspace{-0.15in}
		\end{subfigure}
	\end{center}
	\caption{MNIST: CDF of KL divergence from the output of the unsanitized model to the output of a sanitized model with different C\&W confidences.}
	\label{fig:dkl_carlini}
	\vspace{-0.1in}
\end{figure}

\autoref{fig:dkl_fgsm} compares the CDF of Kullback-Leibler divergence between normal and adversarial examples generated by IGSM after different iterations. It shows that the majority of normal examples have very small divergence, while most adversarial examples have large divergence where more iterations generated examples with higher divergence. \autoref{fig:dkl_carlini} compares the CDF of Kullback-Leibler divergence between normal and adversarial examples generated by the Carlini \& Wagner attack using different confidence levels, which is more prominent.

\autoref{tbl:detection} shows the accuracy of detecting adversarial examples based on the KL divergence from the unsanitized model to a canonical sanitized model. We used a threshold of KL divergence to divide normal and adversarial examples: examples below this threshold were considered normal, and all above adversarial. We determined the threshold by setting a target detection accuracy on normal examples. For example, when we set this target accuracy to 98\%, we needed a threshold of KL divergence of \num{0.0068}. At this threshold, the accuracy of detecting all the Carlini \& Wagner adversarial examples at all the confidence levels is all above 95\%. The accuracy of detecting IGSM adversarial examples is high when the number of iterations is high (e.g., 10 or 15). When the number of iterations is low (e.g., 5), the detection accuracy decreases; however, since the false negative examples have low KL divergence, they are more similar to normal examples and therefore can be classified correctly with high probability as discussed next.

\begin{table*}[t]
	\begin{center}
		\begin{tabular}{llSSSSSSSSSS}
			\toprule
			Attack & & \multicolumn{3}{c}{IGSM (iterations)} & &\multicolumn{6}{c}{Carlini \& Wagner (confidence)}\\ \cline{1-1} \cline{3-5} \cline{7-12}\noalign{\smallskip}
			
			Attack parameter& &5 & 10 & 15 & & 0 & 0.1 & 0.5 & 1 & 3 & 5 \\
			\midrule
			
			Detection accuracy (\%)  & & 33.80 & 85.47 & 96.31 & & 99.26 & 99.26 & 100.00 & 99.26 & 98.52 & 95.56\\
			
			\bottomrule
		\end{tabular}
	\end{center}
	\caption{MNIST: Accuracy of detecting adversarial examples based on the Kullback-Leibler divergence from the unsanitized model to a canonical sanitized model when detection accuracy for normal examples is 98\%.}
	\label{tbl:detection}
	\vspace{-0.35in}
\end{table*}

\begin{figure}[t]
	\centerline{\includegraphics[width=0.6\linewidth]{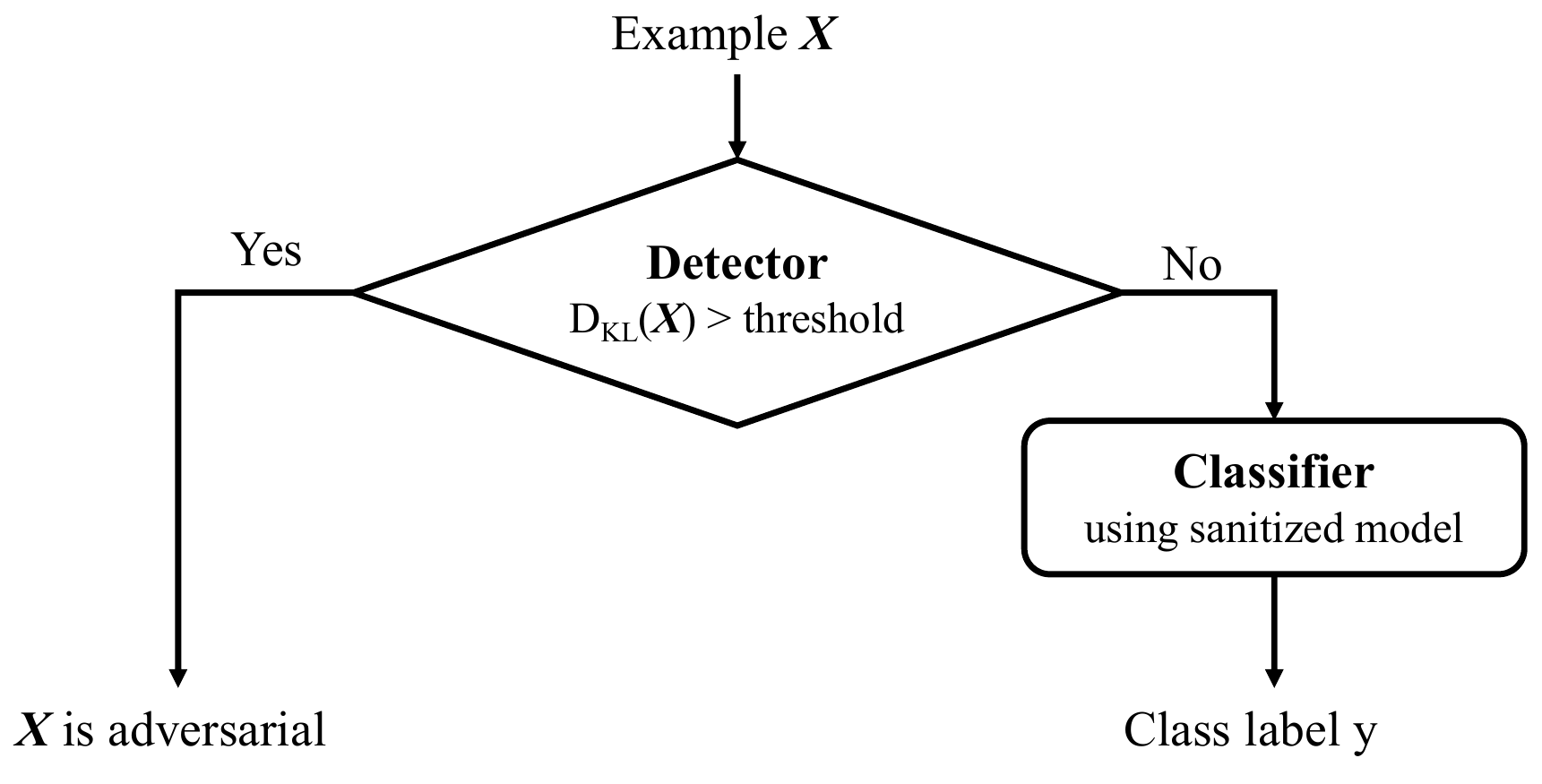}}
	\caption{A framework combining a detector, which detects if an example is adversarial based on the KL divergence from the unsantized model to the sanitized model, and a classifier, which classifies the example using the sanitized model.}
	\label{fig:combined}
	\vspace{-0.1in}
\end{figure}

To take advantage of both the KL divergence for detecting adversarial examples and the sanitized models for classifying examples, we combined them into a framework shown in \autoref{fig:combined}. The framework consists of a detector, which computes the KL divergence from the unsantized model to the sanitized model and rejects the example if its divergence exceeds a threshold, and a classifier, which infers the class of the example using the sanitized model. The framework makes a correct decision on an example when
\begin{itemize}
	\item if the example is regard as normal, the classifier correctly infers its class.
	
	\item if the example is adversarial, the detector decides the example as adversarial \emph{or} the classifier correctly infers its true class.
\end{itemize}

\autoref{tbl:detection_classification} shows the accuracy of this system on adversarial exampled generated by the IGSM attack on a canonical sanitized model on MNIST. At each tested iteration of the IGSM attack, the accuracy of this system on adversarial examples is above 94\%. The accuracy of this system on the normal examples is 94.8\%.

\begin{table*}[t]
	\centering
	\begin{tabular}{cSSS}
		\toprule
		\multirow{2}{*}{IGSM iterations} & \multicolumn{3}{c}{Accuracy (\%)} \\
		& \multicolumn{1}{c}{Detector} & \multicolumn{1}{c}{Classifier} & \multicolumn{1}{c}{Overall} \\
		\midrule
		5 & 33.80 & 99.84 & 99.89 \\
		10 & 85.47 & 72.72 & 96.03\\
		15 & 96.31 & 1.71 & 96.37\\
		\bottomrule
	\end{tabular}
	\vspace{0.1in}
	\caption{Accuracy of the framework in \autoref{fig:combined}. The column ``detection'', ``classifier'', and ``overall'' shows the accuracy of the detector, classifier, and system overall. The overall accuracy on normal examples is 94.8\%.}
	\label{tbl:detection_classification}
	\vspace{-0.35in}
\end{table*}

\subsubsection{SVHN}
\autoref{fig:dkl_carlini_self_svhn} compares the CDF of the Kullback-Leibler divergence of normal examples and adversarial examples generated by the Carlini \& Wagner attack at different confidence levels. We trained the sanitized model with a discard threshold 0.9 (self sanitization). We can see normal examples have small divergence, while all the Carlini \& Wagner adversarial examples under difference confidence levels have large divergence.

\autoref{tbl:svhn_detection} shows the impact of sanitization threshold on the detection accuracy on adversarial examples generated by the Carlini \& Wagner attack. We automatically determined the threshold of KL divergence by setting the detection accuracy on normal examples to 94\%. \autoref{tbl:svhn_detection} shows that as the sanitization threshold increases from 0.7 to 0.9, the detection accuracy increases. However, after the sanitization threshold increases even further, the detection accuracy decreases. This is because after the sanitization threshold exceeds 0.9, the size of the training set decreases rapidly, which causes the model to overfit.

\begin{table*}[t]
	\small
	\begin{center}
		\caption{SVHN: the impact of sanitization threshold on the accuracy of detecting  adversarial examples based on the KL divergence}
		\label{tbl:svhn_detection}
		\begin{tabular}{SSSSSS}
			\toprule
			\multicolumn{1}{c}{Sanitization} &\multicolumn{1}{c}{Training}&\multicolumn{1}{c}{KL divergence}& \multicolumn{3}{c}{Detection accuracy (\%)} \\
			\multicolumn{1}{c}{threshold} &\multicolumn{1}{c}{set size}&\multicolumn{1}{c}{threshold}& \multicolumn{3}{c}{Attack confidence} \\\cline{4-6}\noalign{\smallskip}
			
			& &  & 0 & 0.5 & 1\\
			\midrule
			
			0.7& 39330 & 0.7295 &  93.50 & 94.56 & 94.67\\
			0.8& 38929 & 0.5891  & 93.56 & 94.67 & 95.72\\
			0.9& 38153 & 0.7586  & 94.67 & 94.78 & 95.78\\
			0.99& 34408 & 1.0918  & 90.00 & 91.17 & 92.56\\
			0.999& 28420 & 1.6224 & 83.22 & 85.78 & 87.33\\
			
			\bottomrule
		\end{tabular}
	\end{center}
	\vspace{-0.3in}
\end{table*}

\section{Discussion and future work}

From the observation in \autoref{sec:classification} that the sanitized models will obtain higher robustness, we speculate the causation of this phenomenon is that the outliers will extend the decision boundary and give the model better generalization ability. However, since the outliers are usually not of big proportion and not representative, the extended decision boundary would also include adversarial examples. We call this phenomenon as 'negative generalization'.

The state-of-the-art techniques of handling the negative generalization problem are various advanced adversarial retraining methods, which use adversarial examples as a part of the training data and force the model to correctly classify. These methods are essentially enriching the proportion of the outliers and making the decision boundary more sophisticated to improve the robustness.

In this paper, we focus on another direction. By culling the dataset, we restrict the decision boundary thus also limit the negative generalization on the sanitized model. The sanitized model can help to make a gap of the negative generalization (sensitivity to adversarial examples) between itself and the original model, while the capability of classifying normal examples for both models would stay similar. This lets us leverage the gap to detect adversarial examples.

\autoref{sec:detection} showed that we can use the Kullback-Leibler divergence as a
reliable metric to distinguish between normal and adversarial examples. In our future work, we plan to evaluate if the attacker can generate adversarial examples to evade our detection if she knows our detection method.

\section{Related work}

Most prior work on machine learning security focused on improving the network architecture, training method, or incorporating adversarial examples in training~\cite{Carlini:2017:Adversarial}. By contrast, we focus on culling the training set to remove outliers to improve the model's robustness.

\subsection{Influence of training examples}

Influence functions is a technique from robust statistics to measure the estimator on the value of one of the points in the sample~\cite{Cook:1980,Weisberg:1982}. Koh et al.\ used influence functions as an indicator to track the behavior from the training data to the model’s prediction~\cite{Koh:2017}. By modifying the training data and observing its corresponding prediction, the influence functions can reveal insight of model.  They found that some ambiguous training examples were effective points that led to a low confidence model. Influence Sketching~\cite{Wojnowicz:2016} proposed a new scalable version of Cook’s distance~\cite{Cook:1977,Cook:1979} to prioritize samples in the generalized linear model~\cite{Madsen:2010}. The predictive accuracy changed slightly from 99.47\% to 99.45\% when they deleted 10\%  ambiguous examples from the training set.

\subsection{Influence of test examples}

Xu et al.~\cite{Xu:2018} observed that most features of test examples are unnecessary for prediction, and that superfluous features facilitate adversarial examples. They proposed two methods to reduce the feature space: reducing the color depth of images and using smoothing to reduce the variation among pixels. Their feature squeezing defense successfully detected adversarial examples while maintaining high accuracy on normal examples.

\section{Conclusion}

Adversarial examples remain a challenging problem despite recent progress in defense. In this paper, we study the relationship between outliers in the data set and model robustness and propose a framework for detecting adversarial examples without modifying the original model architecture. We design two methods to detect and remove outliers in the training set and used the remaining examples to train a sanitized model. On both MNIST and SVHN, the sanitized models improved the classification accuracy on adversarial examples generated by the IGSM attack and increased the distortion of adversarial examples generated by the Carlini \& Wagner attack, which indicates that the sanitized model is less sensitive to adversarial examples. Our detection is essentially leveraging the different sensitivity to adversarial examples of the model trained with and without outliers. We found that the Kullback-Leibler divergence from the unsanitized model to the sanitized model can be used to measure this difference and detect adversarial examples reliably.

\section*{Acknowledgment}

This material is based upon work supported by the National Science Foundation under Grant No.\ 1801751.

%
\bibliographystyle{splncs04}
\bibliography{main}

\begin{thebibliography}{10}
\providecommand{\url}[1]{\texttt{#1}}
\providecommand{\urlprefix}{URL }
\providecommand{\doi}[1]{https://doi.org/#1}

\bibitem{Carlini:2017:Adversarial}
Carlini, N., Wagner, D.: Adversarial examples are not easily detected:
  Bypassing ten detection methods. arXiv preprint arXiv:1705.07263  (2017)

\bibitem{Carlini:2017}
Carlini, N., Wagner, D.: Towards evaluating the robustness of neural networks.
  In: {IEEE} Symposium on Security and Privacy (2017)

\bibitem{Cook:1977}
Cook, R.D.: Detection of influential observation in linear regression.
  Technometrics  \textbf{19}(1),  15--18 (1977)

\bibitem{Cook:1979}
Cook, R.D.: Influential observations in linear regression. Journal of the
  American Statistical Association  \textbf{74}(365),  169--174 (1979)

\bibitem{Cook:1980}
Cook, R.D., Weisberg, S.: Characterizations of an empirical influence function
  for detecting influential cases in regression. Technometrics  \textbf{22}(4),
   495--508 (1980)

\bibitem{Goodfellow:2015}
Goodfellow, I.J., Shlens, J., Szegedy, C.: Explaining and harnessing
  adversarial examples. In: International Conference on Learning
  Representations ({ICLR}) (2015)

\bibitem{Koh:2017}
Koh, P.W., Liang, P.: Understanding black-box predictions via influence
  functions. In: International Conference on Machine Learning (2017)

\bibitem{Krizhevsky:2012}
Krizhevsky, A., Sutskever, I., Hinton, G.E.: Imagenet classification with deep
  convolutional neural networks. In: Advances in neural information processing
  systems. pp. 1097--1105 (2012)

\bibitem{Kullback:1951}
Kullback, S., Leibler, R.A.: On information and sufficiency. The annals of
  mathematical statistics  \textbf{22}(1),  79--86 (1951)

\bibitem{Kurakin:2016}
Kurakin, A., Goodfellow, I.J., Bengio, S.: Adversarial examples in the physical
  world. CoRR  \textbf{abs/1607.02533} (2016)

\bibitem{Lecun:1998}
LeCun, Y., Bottou, L., Bengio, Y., Haffner, P.: Gradient-based learning applied
  to document recognition. Proceedings of the IEEE  \textbf{86}(11),
  2278--2324 (1998)

\bibitem{Madsen:2010}
Madsen, H., Thyregod, P.: Introduction to general and generalized linear
  models. CRC Press (2010)

\bibitem{Meng:2017:MagNet}
Meng, D., Chen, H.: {MagNet}: a two-pronged defense against adversarial
  examples. In: ACM Conference on Computer and Communications Security (CCS).
  Dallas, TX (2017)

\bibitem{Metzen:2017}
Metzen, J.H., Genewein, T., Fischer, V., Bischoff, B.: On detecting adversarial
  perturbations. In: International Conference on Learning Representations
  ({ICLR}) (2017)

\bibitem{Moosavi:2016}
Moosavi-Dezfooli, S.M., Fawzi, A., Frossard, P.: Deepfool: a simple and
  accurate method to fool deep neural networks. In: Proceedings of the IEEE
  Conference on Computer Vision and Pattern Recognition. pp. 2574--2582 (2016)

\bibitem{Papernot:2016:Distillation}
Papernot, N., McDaniel, P., Wu, X., Jha, S., Swami, A.: Distillation as a
  defense to adversarial perturbations against deep neural networks. In: {IEEE}
  Symposium on Security and Privacy (2016)

\bibitem{Szegedy:2014}
Szegedy, C., Zaremba, W., Sutskever, I., Bruna, J., Erhan, D., Goodfellow,
  I.J., Fergus, R.: Intriguing properties of neural networks. In: International
  Conference on Learning Representations ({ICLR}) (2014)

\bibitem{Weisberg:1982}
Weisberg, S., Cook, R.D.: Residuals and influence in regression  (1982)

\bibitem{Wojnowicz:2016}
Wojnowicz, M., Cruz, B., Zhao, X., Wallace, B., Wolff, M., Luan, J., Crable,
  C.: “influence sketching”: Finding influential samples in large-scale
  regressions. In: Big Data (Big Data), 2016 IEEE International Conference on.
  pp. 3601--3612. IEEE (2016)

\bibitem{Xu:2018}
Xu, W., Evans, D., Qi, Y.: Feature squeezing: Detecting adversarial examples in
  deep neural networks. In: Network and Distributed Systems Security Symposium
  (NDSS) (2018)

\end{thebibliography}
\end{document}